%% file: pilot_paper_4.0.tex
\documentclass[mathleft
]{an}
\usepackage{graphicx}
\usepackage{times}
\usepackage{natbib}
\bibpunct{(}{)}{;}{a}{}{,}

\input{newcommands}

\begin{document}

\Pagespan{789}{}
\Yearpublication{2006}%
\Yearsubmission{2005}%
\Month{11}%
\Volume{999}%
\Issue{88}%

\title{A pilot study of the radio-continuum emission from MASH planetary
nebulae}

\author{I. S. Boji\v{c}i\'c\inst{1,2}\fnmsep\thanks{Corresponding author:
  \email{ivan.bojicic@mq.edu.au}\newline}
\and  Q. A. Parker\inst{1,3}
\and D. J. Frew\inst{1}
\and  A. E. Vaughan\inst{1}
\and M. D. Filipovi\'c\inst{2}
\and M. L. P. Gunawardhana\inst{4,1}
}
\titlerunning{A pilot study of the radio-continuum emission from MASH PNe}
\authorrunning{I. S. Boji\v{c}i\'c et al.}
\institute{
Department of Physics and Astronomy, Macquarie University, Sydney, NSW 2109, Australia
\and 
University of Western Sydney, Locked Bag 1797, Penrith South DC, NSW 1797,
Australia
\and 
Australian Astronomical Observatory, Epping, NSW 1710, Australia
\and
Sydney Institute for Astronomy, School of Physics, University of Sydney, NSW 2006, Australia 
}

\received{01 Jan 2010}
\accepted{01 Jan 2010}
\publonline{later}


\abstract{
We report an Australia Telescope Compact Array (ATCA) radio-continuum observations of 26 planetary nebulae (PNe) at wavelengths of 3 and 6~cm. This sample of 26 PNe were taken from the Macquarie/AAO/Strasbourg \HA\ PNe (MASH) catalogue and previous lists. We investigate radio detection quality including measured and derived parameters for all detected or marginally detected PNe from this combined sample. Some 11 objects from the observed sample have been successfully detected and parametrized. Except for one, all detected PNe have very low radio surface brightnesses. We use a statistical distance scale method to calculate distances and ionised masses of the detected objects. Nebulae from this sample are found to be large ($>$0.2~pc in diameter) and highly diluted which indicates old age. For  21 PNe from this sample we list integrated \HA\ fluxes and interstellar extinction coefficients, either taken from the literature or derived here from the Balmer decrement and radio to \HA\ ratio methods. Finally, our detected fraction of the MASH pilot sample is relatively low compared to the non-MASH sub-sample. We conclude that future radio surveys of the MASH sample must involve deeper observations with better {\it uv} coverage in order to increase the fraction of detected objects and improve the quality of the derived parameters.}

\maketitle

\section{Introduction}

Planetary nebulae (PNe) are the manifestation of the final evolutionary stage of low and intermediate mass stars (1-8~\msol). For a brief period  \citep[$<5\times10^4$ yr;][]{2010PASA...27..129F} these astrophysical phenomena evolve from compact and dusty, circumstellar envelopes detached from the parent star to the final stage when the  star reaches the cooling path on the HR diagram, the ionised material starts to disperse into the surrounding interstellar medium and the luminosity of the nebula drops swiftly.

A significant number ($\sim$1200) of newly discovered PNe have been listed in the Macquarie/AAO/Strasbourg \HA\ PNe (MASH) PN catalogues \citep{2006MNRAS.373...79P, 2008MNRAS.384..525M} which are a direct product of the Anglo-Australian Observatory/UK Schmidt Telescope (AAO/UKST) \HA\ survey \citep{2005MNRAS.362..689P}.   A significant fraction of MASH PNe represent the oldest stages of PN evolution, dominating the known population at the faint end of the PN luminosity function. 

Based on these discoveries, a pilot observational radio study was conducted in early 2003. This study intended to obtain and explore the radio-continuum data of a relatively small sub-set of the newly discovered PNe. With the known benefits of the radio observational data (e.g. negligible interstellar extinction at cm wavelengths, absolute calibration, and a well studied emission mechanism) the pilot study \linebreak aimed to examine some elementary physical properties of these new PNe, to test observing strategies and to confirm the viability of a large-scale observing program of the large number of MASH discoveries (Boji\v{c}i\'c et al. 2011, in preparation).

In this paper, we present the observational techniques, the radio detection quality, together with measured and derived parameters for the detected objects. We also examined the spectral energy distributions (SEDs) and the correlation between measured radio fluxes and Balmer line fluxes estimated from the SuperCOSMOS H-alpha Survey \citep[SHS;][]{2001PASP..113.1326G} \HA\ images (the MASH part of the sample) or found in the literature (the previously known fraction of the sample). We also present derived distances, radii and physical properties of the detected objects. 

\begin{table*}[t]
\begin{center}
\begin{footnotesize}
\caption[Field names, positions and general properties for PNe in the observed pilot sample.]{Field names, positions and general properties for PNe in the observed pilot sample. The listed coordinates designate the targeted position from the Strasbourg-ESO Catalogue of Galactic Planetary Nebulae  \citep[][A92]{1992secg.book.....A}, \citet[][K02]{2002AN....323..484K} and the MASH I catalogue \citep[][P06]{2006MNRAS.373...79P}. Column (4) is an object status flag as described in \cite{2006MNRAS.373...79P}. Columns (8) and (9) gives the morphological classification and optically determined angular diameter as found in the literature. Column (10) shows detection flags at 6~cm from this project. Objects flagged with ``y'' are positively detected and those with ``n'' not detected.}\label{maintable}
\begin{tabular}{llccclrcccc}
\hline\noalign{\smallskip}
Field	&Name	& PNG &Stat.$^a$&RAJ2000	&DECJ2000& Cat
&Morph$^b$.&$\theta$& det.\\
(1)	&(2)		&(3)	&(4)		&(5)		&(6)	&(7)
&(8)&(9)&(10)\\
\hline\noalign{\smallskip}
mA02&KeWe~2&228.5-11.4&T&06 37 39.1&-18 57 24 		&K02&E/B?&30&n\\
mA12&PHR0724-2021&234.7-02.2&L&07 24 13.1&-20 21 49 	&P06&A&85&n\\
mA08&PHR0724-1757&232.6-01.0&T&07 24 43.4&-17 57 51 	&P06&Rs&169&n\\
mA10&PHR0726-2858&242.5-05.9&T&07 26 04.8&-28 58 23 	&P06&R&32&n\\
\smallskip
wB11&PHR0731-2439&239.3-02.7&L&07 31 59.6&-24 39 04 	&P06&E&19&y\\
mA06&PHR0732-2825&242.6-04.4&T&07 32 17.4&-28 25 18 	&P06&B&27&y\\
wB06&PHR0745-3535&250.3-05.4&T&07 45 41.1&-35 35 04 	&P06&Es&51&n\\
wB07&PHR0755-3346&249.8-02.7&L&07 55 55.5&-33 46 00 	&P06&Ea&100&n\\
wB04&PHR0758-4243&257.8-06.9&T&07 58 26.3&-42 43 53 	&P06&R&25&n\\
\smallskip
mA09&M~3-2&240.3-07.6&T&07 14 49.8&-27 50 23 		&A92&B&11&y\\
wB03&A~23&249.3-05.4&T&07 43 18.0&-34 45 16 			&A92&R&65&y\\
wB10&NGC~2452&243.3-01.0&T&07 47 26.3&-27 20 07 		&A92&A&15&y\\
wB12&M~3-4&241.0+02.3&T&07 55 11.4&-23 38 13 		&A92&-&14&n$^c$\\
wB09&PHR0803-3331&250.4-01.3&T&08 03 12.5&-33 31 02	&P06&B&64&n\\
\smallskip
wB05&KeWe~4&257.8-05.4&T&08 05 33.7&-41 56 51 		&K02&R?&45&n\\
wA08&PHR1833-2632&007.2-08.1&T&18 33 21.5&-26 32 28 	&P06&Rrs&28&n\\
wA02&PHR1835-2751&006.2-09.1&T&18 35 44.6&-27 51 21 	&P06&Eam&22&y\\
wA03&PHR1837-2827&005.9-09.8&T&18 37 54.6&-28 27 31 	&P06&Eas&41&n\\
wA07&PHR1841-2716&007.3-10.1&T&18 41 55.0&-27 16 42 	&P06&Eas&14&n\\
\smallskip
wA09&PHR1848-1829&016.0-07.6&T&18 48 11.3&-18 29 43 	&P06&Ems&19&y\\
wA12&PHR1849-1952&014.8-08.4&T&18 49 24.2&-19 52 14 	&P06&Es&18&y\\
wA04&PHR1852-2749&007.9-12.5&T&18 52 51.5&-27 49 05 	&P06&R&22&n\\
wA10&PHR1857-1750&017.5-09.2&T&18 57 16.8&-17 50 53 	&P06&Eas&11&n\\
wA06&Hf~2-2&005.1-08.9&T&18 32 31.0&-28 43 21 		&A92&-&22&y\\
\smallskip
wA05&He~2-418&004.7-11.8&T&18 44 14.6&-30 19 37 		&A92&E&11&y\\
wA11&A~51&017.6-10.2&T&19 01 01.6&-18 12 13 			&A92&R&59&y\\
\hline
\end{tabular}
\medskip\\
\flushleft
$^a$All PNe from A92 and K02 were designated as true (T).\\
$^b$References for morphological classification of non-MASH PNe: KeWe~2 and KeWe~4: \cite{1998A&AS..130..501K}; M~3-2: \cite{PC98}; A~23 and NGC~2452: \cite{1999A&A...347..169R}; A~51: \cite{1993A&A...279..521S}; He~2-418: \cite{2004MNRAS.353..796R}

$^c$The wB12 field is excluded from the reduction process due to the
insufficient visibility data (see text).

\end{footnotesize}
\end{center}
\end{table*}

\section{Sample selection, observations and data reduction}\label{pilotobsandred}

A group of 17 MASH PNe was randomly selected for observation with the ATCA.  Care was taken that no obvious PN mimics \citep{2010PASA...27..129F, 2010PASA...27..203F} were included in the sample.  The selection criterion, based on angular diameter, was weighted toward objects with \linebreak $\theta_{opt}<$~40\arcsec\ which is approximately the size of the sky-projected shortest baseline in the EW352 ATCA configuration at 6~cm.  A group of 10 known PNe catalogued in \cite{1992secg.book.....A} and  \cite{2002AN....323..484K} was observed along with the selected MASH sample. In order to avoid large interruptions in observation, the main selection criterion for this group was based on their angular proximity to the previously selected MASH sample. 

Selected objects were observed with the ATCA on the 12$^{\textrm{th}}$, 13$^{\textrm{th}}$ and 14$^{\textrm{th}}$ of May 2003. Fields observed on the 12$^{\textrm{th}}$ and 13$^{\textrm{th}}$ are designated in the target listings with a preceding {\bf mA} and {\bf wA}, respectively, and fields observed on the third day (14$^{\textrm{th}}$) has been designated with a preceding {\bf wB}. We present field names, positions, general properties and detection flags (see \S~\ref{sec:detectionquality}) for PNe in the observed pilot sample in Table~\ref{maintable}. Observations were conducted at 6~cm and 3~cm, using the EW352 configuration in snap-shot mode and with a moderate total integration time on each source. The primary calibrator used for the preliminary antennae calibration and for the gain calibration was always PKS~1934-638 with an adopted flux of 2.842~Jy at 6~cm and 5.829~Jy at 3~cm. The summarised observational parameters for this pilot ATCA study are presented in Table~\ref{obs_parameters}.

\begin{table}[t]
\begin{center}
\begin{footnotesize}
\caption{ATCA observational parameters for pilot study. }\label{obs_parameters}
\begin{tabular}{lcc}
\hline\noalign{\smallskip}
Observing wavelength [cm] & 3 & 6\\
Observing frequency [MHz] & 8640 & 4800\\
Bandwidth [MHz]             & 128  & 128\\
ATCA configuration          & EW352&EW352\\
Typical size of the synthesised &&\\
Beam [arcsec]               & 50$\times$20 & 100$\times$40\\
Typical integration time [min] & 35 (15$^a$) & 35 (15$^a$)\\
Typical rms noise [mJy/Beam]& $\sim0.2$ & $\sim0.2$\\
Prim. cal. flux [Jy]: &&\\
PKS~1934-638	& 5.829	& 2.842 \\
Sec. cal. defect:&&\\
0646-306 (mA fields)	& 1\%	& 1\% \\
0736-332 (wA fields)	& 3\%	& 1\% \\
1933-400 (wB fields)	& 1\%	& 1\% \\
\hline\noalign{\smallskip}
\end{tabular}
\medskip\\
\flushleft
$^a$Day three.\\
\end{footnotesize}
\end{center}
\end{table}

\begin{table*}[t]
\begin{footnotesize}
\begin{center}
\caption{Radio parameters of the ATCA radio-detected PNe.}\label{tableflux}
\begin{tabular}{lrrrrrr}
\hline\noalign{\smallskip}
(1)		&(2)	&(3)	&(4)	&(5)	&(6) &(7)\\
Name&$S_{3cm}$&$S_{6cm}$&$S_{NVSS}$&$\theta_{3cm}$&$\theta_{6cm}$&$\theta_{opt}
$\\
		&[mJy]&[mJy]&[mJy]& [arcsec]&[arcsec]&[arcsec]\\
\hline\noalign{\smallskip}
PHR0731-2439	&3.5$\pm$0.4	&3.2$\pm$0.4
&4.0$\pm$0.7&14$\pm$3&15$\pm$3&19\\
PHR0732-2825	&-		&0.8$\pm$0.2	&-&-& - &27\\
M 3-2		&2.8$\pm$0.3	&2.3$\pm$0.3	&2.6$\pm$0.5&16$\pm$4&-&11\\
A 23		&5.1$\pm$0.5	&5.1$\pm$0.6	&4.2$\pm$0.6&33$\pm$7&-&65\\
NGC 2452	&47.0$\pm$4.7	&52.9$\pm$5.3
&56.0$\pm$1.7&15$\pm$3&13$\pm$3&15\\
PHR1835-2751	&1.7$\pm$0.9	&1.8$\pm$0.3	&-&13$\pm$4&-&22\\
PHR1848-1829	&2.3$\pm$1.3	&1.8$\pm$0.2	&3.0$\pm$0.6&8$\pm$3&-&19\\
PHR1849-1952	&2.9$\pm$0.4	&3.3$\pm$0.4	&-&23$\pm$6&21$\pm$5&18\\
Hf 2-2		&4.7$\pm$0.5	&3.2$\pm$0.9	&4.4$\pm$0.5&23$\pm$5&-&22\\
He 2-418	&2.3$\pm$0.9	&3.2$\pm$0.4	&3.1$\pm$0.5&-&-&11\\
A 51		&7.1$\pm$3.1	&9.2$\pm$1.0
&7.3$\pm$1.2&62$\pm$15&69$\pm$15&59\\
\hline
\end{tabular}
\end{center}
\end{footnotesize}
\end{table*}

Unfortunately, for all of the observed objects, data from antennae CA01 and CA02 (the shortest baseline) have been corrupted at the beginning and at the end of the observing runs because of the shadowing effect\footnote{For a full explanation of the shadowing effect refer to \cite{2004MNRAS.349.1365S}.} . Corrupted visibilities were flagged-out prior to the calibration process. After the initial flagging process the data-lost was approximately 15-40 per cent of the total integration time in the shortest baseline and a significant part of correlations with the CA01 antenna. The wB10 field is left with only one $uv$ cut in the shortest baseline visibility data, and the wB12 field with no shortest baseline visibility data at all and with only two $uv$ cuts in the longer baselines. The wB10 field fortunately contained a strong and relatively compact ($\theta<20$\arcsec) program source. We argue that the loss of large structure information should not have a crucial effect on the final result. Due to the insufficient visibility data, the wB12 field is completely excluded from further analysis. In the wB02 field strong interference in the CA01-CA02 correlation at 3~cm was noticed (unrelated to the shadowing effect) . Since this interference could not be properly flagged, this part of the data set was also excluded from further processing. The program source A 23, observed in the wB02 field, has an optically determined angular size of 65\arcsec. Thus, we anticipated that the 3~cm flux density could be underestimated due to a lack of the flagged correlations (short baseline) data.

Further processing (calibration, deconvolution and parameterisation) was performed using standard \miriad\ procedures \citep{miriad} with no attempt to employ a self-calibration due to low signal to noise ratio in the observed fields. The ``dirty'' images were created using the natural weighting scheme and excluding antenna 6 (at 6~km). 

Detected objects were parameterised using the IMSAD task. Measured integrated flux densities are tabulated in Table~\ref{tableflux} (columns 2 and 3). If the IMSAD task reported a proper deconvolution the angular diameters were calculated using the Gaussian deconvolution method as described in \cite{2000MNRAS.314...99V}. The derived angular diameters from 6~cm and 3~cm maps are presented in columns (5) and (6) respectively. We note that because of the small declinations and incomplete (and non uniform) $uv$ coverage the synthesised beams have strong eccentricities ranging from 0.87 to 0.97. Therefore, we used only a minor deconvolved diameter for the calculation of angular diameters.

We expected the influence of several systematic effects to affect the accuracy of derived parameters due to the limited integration time per field, a considerable loss of the short baseline information, the strong elongation of the synthesised beam and the intrinsic low brightness of observed objects. This pilot study gave us an excellent opportunity to measure the observational limits of MASH PN detectability, the extent of uncertainties and a chance to examine the observational components which have to be improved in the following full-scale experiment.

\section{Detection quality}\label{sec:detectionquality}

\begin{figure*}[t]
\begin{center}
\includegraphics[scale=0.245]{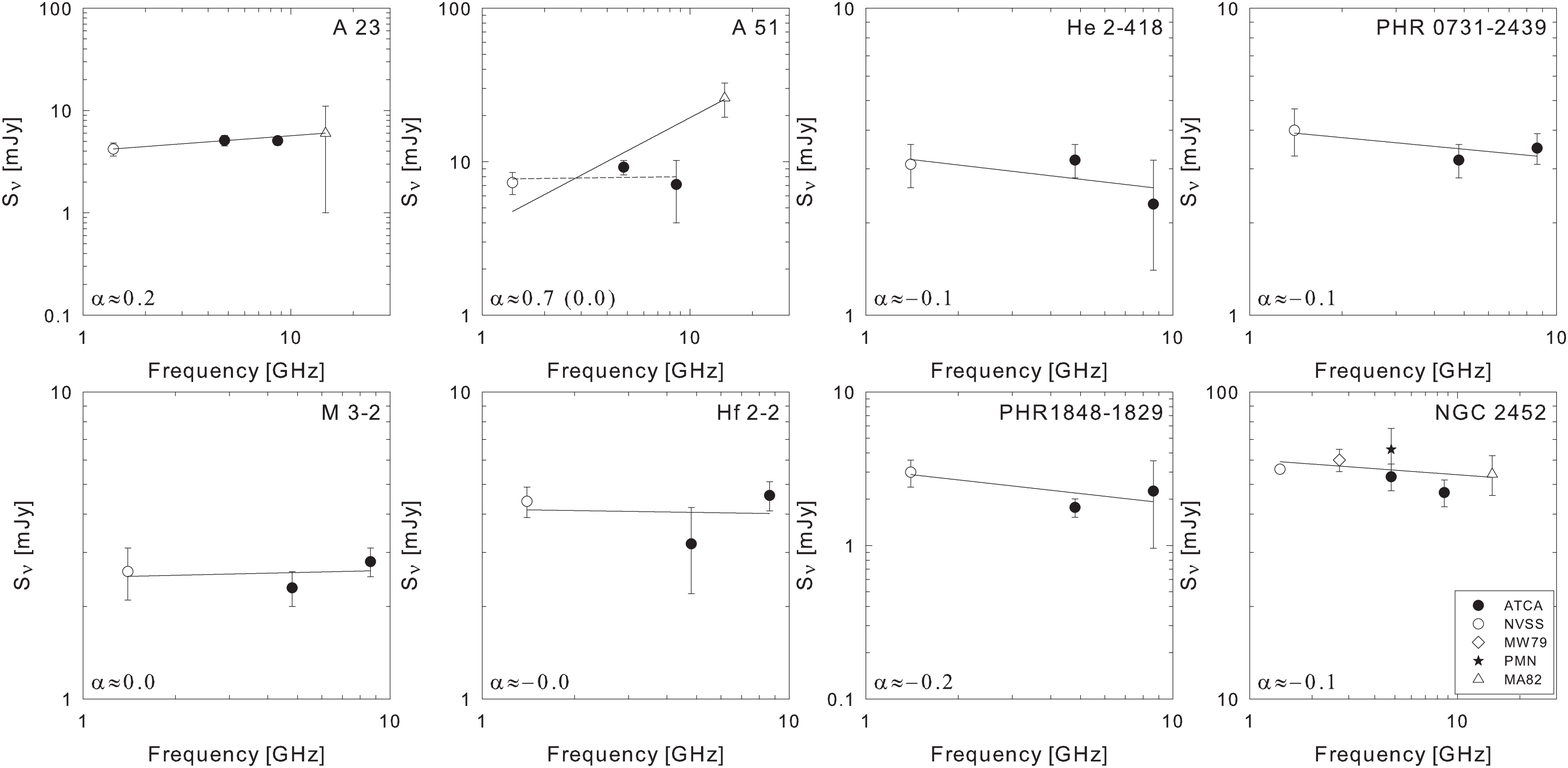}
\caption[radio-continuum spectral distribution plots of known PNe from the pilot sample]{The radio-continuum SED plots of PNe from the pilot sample. Filled circles represent the integrated flux densities measured from the new ATCA radio images (this paper). Other values are from: MW79 \citep{1979A&AS...36..169M}; MA82 \citep{1982A&AS...50..209M}; PMN~ \citep{1994ApJS...90..179G}; NVSS \citep{1998AJ....115.1693C}. The straight line represent the best fit of the power-law function ($S(\nu)\propto\nu^{\alpha}$) to all data-points from the obtained empirical distribution. In the case of PN A~51 the power law function is also fitted to a sub-set of flux densities (ATCA and NVSS; dashed line). The value of a spectral index obtained from the best fit is given in the lower left corner.}
 \label{spindexs:1}
\end{center}
\end{figure*}

We detected 11 of the 26 observed objects in our radio maps (25 if we exclude the wB12 field which was not fully processed). Six detected objects are associated with PNe positions from the \cite{1992secg.book.....A} and \cite{2002AN....323..484K} catalogues and five from the MASH~I catalogue, with one object from the MASH catalogue positively detected only at 6~cm. That give us a detection quality strongly in favour of known PNe with around 60\% detected, against 30\% of detected MASH PNe. This two-to-one detection ratio between the known and MASH samples will not change significantly even if we exclude three PNe with optical sizes significantly larger than the synthesised beam (PHR0724-2021,  PHR0724-1757, and PHR0755-3346).

The only object which is detected at only one frequency (6~cm) is PHR0732-2825. The 6~cm detection is at the 5~$\sigma$ threshold level and it is not a surprise that no emission is visible at 3~cm. Even though possible non-thermal emission could produce such an effect, due to its distinct bipolar morphology we have no doubt that the object in question is a genuine PN and that the missing 3~cm detection is solely produced by the insufficiently sensitive observation.

Assuming that the intrinsic properties of this preliminary MASH sample conform to the general characteristics of the full MASH sample it is clear that it can be expected that most MASH PNe would be below or at the edge of sensitivity of a similar quick ATCA survey (of 1~mJy zero level noise). Obviously, faint and extended ($\theta>$100\arcsec) PNe are not detectable using this observational configuration. Even with much larger integration times, the measured fluxes will be strongly affected by the missing flux effect. 

\section{Radio-continuum spectral energy distribution}\label{spindex_dist}

With a presumption of free-free emission as the primary radio emission mechanism in PNe \citep{1984ASSL..107.....P,1989agna.book.....O}, a construction of the SED plots can be used as a crude test of the reliability of our results.

First, a thorough literature search for published radio data of the selected sample was undertaken. We collected all radio data which can be used to cross-correlate with the obtained parameters. We found flux densities at 6~cm for M~3-2, A~23, NGC~2452 and A~51 catalogued in \cite{1992A&AS...94..399C}. However, the only genuine 6~cm flux density observation is for NGC~2452 (55$\pm$9~mJy) originating from \cite{1975A&A....38..183M}. The catalogued values for A~23 and A~51 were calculated from the 14.7~GHz flux reported in \cite{1982A&AS...50..209M} and for M~3-2 from the \HB\ flux from \cite{1990ApJ...359..392K}. We also found a 5~GHz flux density (65$\pm$11~mJy) for NGC~2452 detected in the PMN Tropical survey \citep{1994ApJS...90..179G}, which is in excellent agreement with the \cite{1975A&A....38..183M} value. From \cite{1982A&AS...50..209M} we found flux measurements at 14.7~GHz for A~23 (6$\pm$5~mJy), A~51 (26$\pm$5~mJy) and NGC~2452 (54$\pm$5~mJy). At 0.408~GHz we found flux upper limit of 60$\pm$25~mJy \citep{1982MNRAS.199..141C} and at 2.7~GHz we found 60$\pm$10~mJy \citep{1979A&AS...36..169M} for NGC~2452. Finally, the NRAO VLA Sky Survey \citep[NVSS:][]{1998AJ....115.1693C} covers this part of the sky at 1.4~GHz. Flux densities from NVSS, for 8 detected PN, are tabulated in column (4) in Table~\ref{tableflux}. Published data, together with the data from our observations, were used for the construction of SEDs for 6 known and 2 MASH PNe (Fig.~\ref{spindexs:1}). All plots were produced only for objects where the independent data were available. 

Except in the case of the A~51, our measurements are in a good agreement with that expected for optically thin nebulae. The spectra observed in A~51 appear to have a signature of two component emission or some additional emission mechanism at frequencies above 10~GHz. We examined this PN in more detail in section~\ref{individual_objs}.

\begin{table*}[t]
\begin{footnotesize}
\begin{center}
\caption{Analysis of the extinction coefficients calculated from different
methods.}\label{ext1}
\begin{tabular}{llcccc}
\hline\noalign{\smallskip}
Name&		log F(\HA)&
$E(B-V)$&$E(B-V)_{6cm}$&$E(B-V)_{20cm}$\\
\hline\noalign{\smallskip}
M~3-2		&-12.11$^{(1)}$ 	&0.56$^{(2)}$				&0.43	&		0.43	\\
He~2-418		&-12.05$^{(3)}$ 	&0.42$^{(1)}$				&0.52	&		0.45	\\
Hf~2-2		&-11.63$^{(4)}$ 	&0.50$^{(5)}$				&0.10	&		0.18	\\
A~51		&-11.35$^{(4)}$ 	&0.12$^{(6)}$ (0.22)$^{(7)}$	&0.27	&		0.12	\\
A~23		&-11.88$^{(4)}$ 	&1.69$^{(3)}$ (0.65)$^{(7)}$	&0.55	&		0.42	\\
KeWe~4	&-12.92$^{(8)}$ 	&0.70$^{(8)}$				&-		&		-	\\
NGC~2452	&-10.81$^{(1)}$ 	&0.36$^{(9)}$				&0.56	&		0.53	\\
\hline\noalign{\smallskip}
\smallskip
Name&		log F(\HA)$^{(10)}$&	
$E(B-V)^{(11)}$&$E(B-V)_{6cm}$&$E(B-V)_{20cm}$\\
\hline\noalign{\smallskip}
PHR0724-2021&	-12.47&	-			&	-	&	-\\
PHR0732-2825&	-12.67&	0.51			&	0.53	&	-\\
PHR0726-2858&	-12.76&	0.26			&	-	&	-\\
PHR1835-2751&	-13.05&	-			&	1.26	&	-\\
PHR1837-2827&	-13.18&	-			&	-	&	-\\
PHR1841-2716&	-13.72&	-			&	-	&	-\\
PHR1848-1829&	-12.67&	-			&	0.88	&	1.05\\
PHR1849-1952&	-12.50&	-			&	0.98	&	-\\
PHR1857-1750&	-13.22&	-			&	-	&	-\\
PHR0758-4243&	-13.50&	0.77			&	-	&	-\\
PHR0745-3535&	-12.38&	-			&	-	&	-\\
PHR0755-3346&	-12.07&	0.65			&	-	&	-\\
PHR0803-3331&	-12.32&	0.50			&	-	&	-\\
PHR0731-2439&	-12.53&	1.19			&	1.00	&	1.03\\
\hline\noalign{\smallskip}
\end{tabular}
\medskip\\
\flushleft
References: 
$^{(1)}$\cite{1989ApJS...69..495S}; 
$^{(2)}$\cite{1991A&A...248..197K}; 
$^{(3)}$\cite{2004MNRAS.353..796R};
$^{(4)}$ Frew et al. (2011, in preparation);
$^{(5)}$\cite{1991A&AS...89..237A};
$^{(6)}$\cite{1983ApJ...271..188K}; 
$^{(7)}$ \cite{2008PhDT.......109F};
$^{(8)}$\cite{1998A&AS..130..501K}; 
$^{(9)}$\cite{1998A&A...337..866P};
$^{(10)}$ Gunawardahana et al. (2011, in preparation);
$^{(11)}$this work.\\
\end{center}
\end{footnotesize}
\end{table*}

\section{\HA\ surface brightness and interstellar extinction}\label{halphabright}

We found published \HA\ and/or \HB\ fluxes and extinction coefficients for seven known PNe from this sample. All but one of these (KeWe~4) have been positively detected in our total intensity radio images.  For the MASH part of the sample the integrated \HA\ fluxes were obtained from intensity calibrated SHS images.  A set of integrated \HA\ fluxes for 14 MASH PNe were taken from Gunawardahana et al. (PASA, in preparation). The integrated fluxes, not corrected for reddening, are presented in Table~\ref{ext1}. The colour excess $E(B-V)$ is estimated for five objects from our MASH sample with available flux-calibrated spectra using the standard \linebreak  Balmer decrement method.  The flux in the Balmer lines (\HA\ and \HB) have been fitted with Gaussians using the IRAF\footnote{IRAF is distributed by the National Optical Astronomy Observatories, which is operated by the Association of Universities for Research in Astronomy, Inc. (AURA) under cooperative agreement with the National Science Foundation} task {\sc splot}. For PHR0755-3346 the value for $E(B-V)$ is calculated from the $c_{H\beta}$ tabulated in \cite{2008PhDT.......109F}. 

It is expected that radio-continuum and Balmer lines emission will be well correlated, due to the same dependence on the nebular density, except in the case of optically thick radio-continuum emission.  In the range of used frequencies (1-10~GHz), and assuming canonical electron temperature ($T_e\cong10^4$K) and density ratios ($n(He^+)/n(H^+)$ = 0.11, and $n(He^{2+})/n(H^+)$ = 0.013 the $E(B-V)$ can be calculated from the measured radio flux $F_{\nu}$ at frequency $\nu$ and the measured and reddening-corrected \HA\ flux using \citep{1984ASSL..107.....P}:

{\setlength{\mathindent}{0pt}
\begin{equation}\label{raditohalpha}
\biggl(\frac{F_{\nu}}{\textrm{mJy}}\biggr)=
\biggl(\frac{F(H\alpha)}{\textrm{erg~cm}^{-2}~\textrm{s}^{-1}}\biggr)\
1.02\times10^{10}
\MTelc^{0.53}\nu^{-0.1}
\end{equation}}

\noindent where the standard theoretical ratio of \HA/\HB=2.85 is assumed. We list the extinction coefficients found from this method in the last two columns of Table~\ref{ext1}.

\begin{table*}[t]
\begin{footnotesize}
\begin{center}
\caption{Derived parameters.}\label{tab:physpropt1}
\begin{tabular}{lrrrrrrrr}
\hline\noalign{\smallskip}
Name & $log{(T_b)}$&D$_{P04}$&D$_{S08}$&D$_{F08}$&D$_{adopt}$&R&$n_e$    
&M$_{ion}$\\
     && (kpc) &(kpc) &(kpc)&(kpc) &(pc)   &(cm$^{-3})$&(\msol)\\
\hline\noalign{\smallskip}
PHR0726-2858	&-		&-	&-	&8.8		&8.8		&0.68		&$<$40	&$<$0.57\\
PHR0731-2439	&-0.2	&5.4	&5.0	&5.4		&5.3		&0.24		&220		&0.14\\
PHR0732-2825	&-1.1	&6.9	&5.4	&7.7		&6.7		&0.44		&60		&0.23\\
PHR0755-3346	&-		&-	&-	&1.9		&1.9		&0.46		&$<$20	&$<$0.09\\
\smallskip
PHR0758-4243	&-		&-	&-	&11.4	&11.4	&0.69		&$<$40	&$<$0.59\\
PHR0803-3331	&-		&-	&-	&4.2		&4.2		&0.65		&$<$20	&$<$0.25\\
PHR1835-2751	&-0.6	&6.0	&5.2	&-		&5.6		&0.30		&130		&0.16\\
PHR1848-1829	&-0.4	&6.4	&5.7	&-		&6.1		&0.28		&160		&0.16\\
PHR1849-1952	&-0.1	&5.5	&5.2	&-		&5.4		&0.24		&240		&0.14\\
\smallskip
M~3-2		&0.2		&7.5	&7.4	&8.0		&7.6		&0.20		&350		&0.13\\
He~2-418		&0.3		&6.8	&7.0	&8.3		&7.4		&0.20		&420		&0.14\\
Hf 2-2		&-0.3	&5.1	&4.6	&4.5		&4.7		&0.25		&180		&0.13\\
A 51			&-0.7	&2.4	&2.1	&1.9		&2.1		&0.30		&110		&0.13\\
A 23			&-1.0	&2.8	&2.2	&2.9		&2.6		&0.41		&60		&0.18\\
NGC2452		&1.3		&2.7	&3.3	&3.4		&3.1		&0.11		&1640	&0.11\\
\hline
\end{tabular}
\end{center}
\end{footnotesize}
\end{table*}

\section{Distances and Ionised Masses}\label{distmass}

Further parameterisation of the ATCA detected objects requires knowledge of reliable distances. For four previously known PNe (M~3-2, A~51, A~23 and NGC~2452) distances, calculated using some of the statistical distance scale \linebreak method, have been published in \cite{1992A&AS...94..399C,1998A&AS..130..501K,2004MNRAS.353..589P}. 

In this study, statistical distance scale methods from \linebreak  \cite{2004MNRAS.353..589P}, \cite{2008ApJ...689..194S} and \cite[][F08]{2008PhDT.......109F} has been used to calculate distances to the observed PNe. For the F08 scale we calculated the \HA\ surface brightness from the integrated \HA\ fluxes corrected for reddening. Optically determined angular sizes are used throughout our calculations as they were considered more reliable. The derived distances ($D_P$/kpc) and radii ($R_P$/pc) are tabulated in Table~\ref{tab:physpropt1}.

Considering an optically thin, ionised region of gas with approximation of constant electron temperature ($T_e$/K), the electron density ($n_e/$cm$^{-3}$) and ionised mass ($M_{ion}/$\msol) can be calculated from \citep{1987A&AS...71..245G}:

{\setlength{\mathindent}{0pt}
\begin{equation}
\Mnelc=\zeta\times S_{6cm}^{\frac{1}{2}}T_{e}^{\frac{1}{4}}D^{-\frac{1}{2}}
\theta^{-\frac{3}{2}}\varepsilon^{-\frac{1}{2}}\times
\biggl(\frac{1+y+3xy}{1+y+xy}\biggr)^{-\frac{1}{2}},
\end{equation}}

\noindent where $\zeta=4.96\times10^{2}$ and:

\begin{equation}
\MMion=1.18\times10^{-8}\Mnelc D^3\theta^3\varepsilon\frac{1+4y}{1+y+xy},
\end{equation}

\noindent where $S_{6cm}$ is the 6~cm flux density (mJy), D is the distance (kpc), $\theta$ is angular diameter (arcsec) of the source, $y$ is the He abundance, $x$ is the fraction of \HEii\ and $\varepsilon$ is the volume filling factor defined as the ratio of the electron density (\nelc) averaged over the volume to the electron density averaged over the mass \citep{1982ApJ...260..612D}. We assumed an electron temperature of $10^4$~K, He/H ratio of 0.11, average filling factor of 0.35 \citep{1994A&A...284..248B}, and that half of the He atoms are doubly ionised. The distance adopted for calculation of \nelc\ and \Mion\ is the average of all obtained values except when one of the used methods shows a clear deviation from the mean value (in which case we used an average of the similar diameters). The estimated \nelc\ and \Mion\ are tabulated in Table~\ref{tab:physpropt1}.

We compare derived parameters with parameters for the well known Helix Nebula (NGC 7293). The Helix Nebula is one of the best studied large, nearby and low surface brightness PN with a reliable distance estimate, based on the trigonometric parallaxes method, of only 216~pc \linebreak \citep{2009AJ....138.1969B}. The angular radii (seen in the radio-continuum) is $\sim$5~arcmin implying the radius of the strongly ionised gas of about 0.35~pc. The radio flux at 5~GHz reported in \cite{1975A&A....38..183M} of 1.292~Jy is very likely an upper limit for the real flux considering the influence of several background sources. Following the same method of \cite{1987A&AS...71..245G}, and using 0.75 for the filling factor, \linebreak \cite{1999ApJ...522..387Y} determined the ionised mass of the nebula to be $\sim0.36$~\msol. However, \cite{2002ApJ...574..179R} argue that the filling factor approximation of $\varepsilon=0.75$ is too large for the Helix and that a more realistic value for the ionised mass is $\sim0.074$~\msol\ (with $\varepsilon=0.005$) given by \cite{1994A&A...284..248B}. The electron density in this nebula varies from 30 to 120 cm$^{-3}$ \citep{1999ApJ...517..782H} with a mean value in the main torus of about 60~cm$^{-3}$ \citep{1998AJ....116.1346O}. Scaling down the flux at 5~GHz and the angular diameter of 10~arcmin to the mean value of distances to our detected objects of 5~kpc (see Table~\ref{tab:physpropt1}) will give $S_{6cm}=2.5$~mJy and angular diameter of $\theta=30$~arcsec. Using the same filling factor as used in our sample ($\varepsilon=0.35$) we calculate an electron density and ionised mass of \nelc$\approx120$~cm$^{-3}$ and $M_{ion}\approx0.17$~\msol\ respectively. Mean values for electron density, ionised mass and radius found from the detected portion of the pilot sample (excluding   NGC~2452) are 200~cm$^{-3}$, 0.15~\msol\ and 0.3~pc, respectively. Clearly, the average physical properties of PNe from our pilot sample are very similar to those of the Helix Nebula. 

\section{Notes on individual objects}\label{individual_objs}

In this section, we list individual notes on observed objects for which measured, derived or previously catalogued parameters are ambiguous or in disagreement.


\subsection{Hf~2-2 (PNG005.1-08.9)}
Hf~2-2 is a PN with a variable, close binary central star \citep{2008AJ....136..323D,2010PASP..122..524L}. \cite{2006MNRAS.368.1959L} presented an extreme abundance discrepancy factor (ADF) of about 70 for this nebula  (the typical ADF is $\sim2$ for most PNe). The ADF is a quantification of the abundance discrepancy, found in PNe and HII regions,  between abundances determined from optical recombination lines and those \linebreak found from collisionally excited lines \citep{2006IAUS..234..219L}.  A possible explanation for this discrepancy could be strong temperature gradients within the nebula \citep{1967ApJ...150..825P}, or inner, hydrogen deficient clumps embedded in the diffuse nebula \citep{2009ApJ...695..488Z}. The electron density determined from the hydrogen recombination spectrum near the Balmer jump region of $\Mnelc\approx$400~cm$^{-3}$ \citep{2004MNRAS.351..935Z} is a factor of two larger than our radio-continuum determined value.

\subsection{A~23 (PNG249.3-05.4)} 
This is a moderate excitation \citep[EC~$=5$;][]{1999A&A...347..169R}, round PN with a distinct thin shell \citep{1992secg.book.....A}. \linebreak \cite{1999A&A...347..169R} measured strong [\OIII] lines and did not detect \HEii\ indicating that it might be only moderately optically thin. A strong divergence from the radio flux expected from the measured \HB\ and constructed SED for this nebula imply the possibility of mild self-absorption effects in the radio-continuum. Due to its large angular size and the small integration time (only three 5~min cuts or 15~min of total integration time) we suggest that the integrated flux, reported here, should be taken as a lower limit. Another indicator that this object is not properly sampled is that the Gaussian fitting failed at 6~cm, while the estimated angular diameter at 3~cm is twice as small as the one determined from optical observation. On the other hand, the extinction coefficient estimated from the Balmer line ratio \citep{1999A&A...347..169R} could be overestimated due to the possible internal absorption. Also, a deviation from the optical diameter could suggest that the majority of the measured radio emission is actually produced in some smaller structure. In order to positively distinguish between possible situations high resolution and high sensitivity radio observations are needed.

\subsection{NGC~2452 (PNG243.3-01.0)} 
A PN ionised by a faint but extremely hot ($\sim140\times10^4$~K) Wolf-Rayet central star \citep{2001A&A...367..983P}. The nebula consists of high gas density knots embedded in a diffuse body. NGC~2452 is the brightest radio object in this sample and with most independent flux measurements. Our 5~GHz flux appear to be in better agreement with the predicted optically thin SED than measurements from the PMN survey. As stated above, the excluded shortest baselines at 3~cm imply that the integrated flux from this wavelength must be considered as a lower limit. The observed flat SED down to $\sim1$~GHz is in contradiction to an extremely high electron density of $\sim40\times10^4$~cm$^{-3}$ found by \cite{1999ApJ...525..863F} from the [\NeIV] diagnostic. On the other hand electron density determined from the [\SII] doublet diagnostic \linebreak (\nelc$\approx1.5\times10^3$~cm$^{-3}$) from \cite{1998A&A...337..866P} are in excellent agreement with our estimate ($1.64\times10^3$~cm$^{-3}$).

\begin{figure}[t]
 \begin{center}
\includegraphics[scale=0.4]{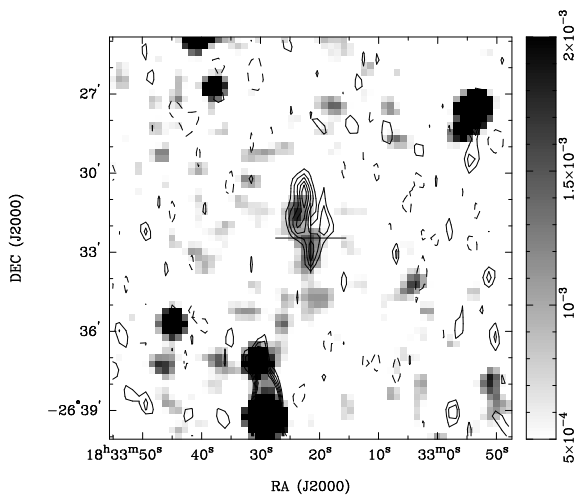}
\caption[The NVSS total intensity map centered on the position of possible detected
MASH PN PHR1833-2632.]{NVSS total intensity map centered on the position of
possible detected PN PHR1833-2632. Overlaid are contours from our 6~cm image.
The contour levels are: -2, 2, 3, 4, 5 and 6 $\times$ 0.1~mJy~beam$^{-1}$. The
sidebar quantifies the pixel map and its units are Jy~beam$^{-1}$.}
 \label{phr1833nvss}
 \end{center}
\end{figure}

\subsection{PHR1833-2632 (PNG007.2-08.1)} 
This MASH PN has an optical diameter of 28\arcsec\ but our radio image shows extended structure above 2$\sigma$ larger than 1\arcmin\ and a radio peak above 5$\sigma$ placed $\sim$100\arcsec\ from the centroid of the nebula. In order to determine if this extended structure is real or a possible artefact from a strong source placed $\approx$10\arcmin\ from the field centre, the 6~cm contour image is compared with the 1.4~GHz NVSS total intensity map. As can be seen the extended structure in Fig.~\ref{phr1833nvss}, which is visible in our 6~cm map, is correlated with a similar extended emission which is at the edge of the adopted zero level flux for the NVSS. This gives confidence that the observed emission is real. However, the question if this emission is actually correlated to the much smaller PN still stands open.

\subsection{Abell~51 (PNG017.6-10.2)}  
The flux density found from a 14.7~GHz Parkes observation \citep{1982A&AS...50..209M} is almost three times larger than both the NVSS and our ATCA flux densities. The spectral energy distribution, constructed from all four measurements imply that the bulk of the radio emission, up to 14.7~GHz, is coming from an optically thick environment. However, the calculated brightness temperature at 1.4~GHz of $T_b\approx2$~K and excellent agreement with the flux predicted from \HB\ (from \cite{1983ApJ...271..188K}) imply optically thin free-free emission at frequencies $\nu>1$~GHz. One can argue that both NVSS and ATCA flux measurements could be affected by the missing flux problem due to the relatively large angular size of this PN ($\theta_{opt}\approx60$\arcsec). However, our radio determined angular diameters, at both 3 and 6~cm (62$\pm$15\arcsec and 69$\pm$15\arcsec, respectively), are in excellent agreement with the optical value which strongly imply that ATCA properly sampled this PN. Also, as stated in \cite{1998AJ....115.1693C}, the VLA, in the used configurations, starts to be insensitive for structures larger than several arcminutes. This puts A~51 well below the NVSS filtering limit. On the other hand, the Parkes beamwidth at this frequency is $\sim$2\arcmin. Except for a few ambiguous field detections of order of 1-2~mJy we did not found any additional and strong source in the $\sim$5\arcmin\ and $\sim$3\arcmin\ (at 6~cm and 3~cm respectively) radii from the observed PN. This allows us to conclude that the measured 14.7~GHz, if not affected by some observational and/or systematic effect, flux should originate solely from this object. Consequently we believe that the previously published 14.7~GHz flux density of 26~mJy is greatly overestimated and that the more realistic value for the flux density at this frequency (roughly estimated from the constructed, optically thin, \linebreak radio-continuum SED) is $\sim6$~mJy. A deeper study of A~51 is required to resolve its intrinsic density structure.

\section{Summary}

Using the ATCA radio telescope a representative sample of faint and extended Galactic PNe have been observed. The sample consists of 10 previously known and 17 newly catalogued PNe from the MASH~I catalogue. Some 11 objects from the observed sample have been successfully detected and parameterised. For nine partially resolved PNe we determined radio angular diameters. For six PNe these are in good agreement with optically determined values. 

We examined SEDs in the cm range for eight radio-detected PNe from our sample for which we found other independent radio-continuum observations. All PNe from this sub-sample appear to emit in the radio optically thin regime which imply a more diffuse ionised medium is present. 

Except for one object, all detected PNe have very low radio surface brightness. We use several statistical distance scale methods to calculate distances, electron densities and ionised masses for the detected PNe. Except for NGC~2452, all of PNe from this sample are found to be moderately large ($>$0.2~pc in physical diameter) and moderately to highly diluted (with electron densities in range of 60~cm$^{-3}$ to \linebreak 1640~cm$^{-3}$ with a median of 180~cm$^{-3}$). The ionised \linebreak masses are found to be concentrated in a relatively narrow range around 0.15~\msol. 

The results presented, together with a review of previous radio detections of MASH PNe \citep{2011MNRAS.412..223B}, give us an  insight into the general radio-continuum properties of this new MASH sample. From the derived physical properties we can see that detected MASH PNe are not significantly different from the known PNe sub-sample. However, it appears that, unsurprisingly, MASH PNe are placed at the faint end of the radio luminosity distribution.

The detection rate of $30\%$ is rather poor. All future radio surveys of MASH PNe should involve significantly deeper observations with better $uv$ coverage in order to improve the detection rate and quality of the derived parameters. Very careful planning of observations is necessary to reconcile limited observational time and the size of the sample which can properly represent the full catalogue. Using  experience gained from this pilot study, we examined the methods and techniques which would improve our future observations. These upcoming observations will form the core of our new detailed investigation of the radio-continuum properties of MASH PNe (Boji\v{c}i\'c et al. in preparation).

\end{document}

%% file: newcommands.tex
\newcommand{\NeIV}{Ne\,{\sc iv}}
\newcommand{\SII}{S\,{\sc ii}}
\newcommand{\HA}{H{$\alpha$}}
\newcommand{\OIII}{O\,{\sc iii}}

\newcommand{\HB}{H{$\beta$}}

\newcommand{\HEii}{He\,{\sc ii}}
\newcommand{\msol}{M$_\odot$}

\newcommand{\miriad}{{\sc miriad}}

\newcommand{\nelc}{{$n_e$}}

\newcommand{\Mnelc}{n_e}
\newcommand{\MTelc}{\textrm{T}_e}

\newcommand{\Mion}{M$_{ion}$}
\newcommand{\MMion}{\textrm{M}_{ion}}